\documentclass[11pt]{article}
\usepackage{amsmath}
\usepackage{amsfonts}
\usepackage{amssymb}
\usepackage{graphicx}
\usepackage{bm}
\usepackage{color}
\usepackage{amscd}

\def\bea{\begin{eqnarray}}
\def\eea{\end{eqnarray}}

\begin{document}
\begin{center}
\LARGE {\bf  Black hole conserved charges in Generalized Minimal Massive Gravity}
\end{center}
\begin{center}
{\bf M. R. Setare\footnote{rezakord@ipm.ir} }\hspace{1mm} ,
H. Adami \footnote{hamed.adami@yahoo.com}\hspace{1.5mm} \\
 {Department of Science, Campus of Bijar, University of  Kurdistan  \\
Bijar, IRAN.}
 \\
 \end{center}
\vskip 3cm

\begin{abstract}
In this paper we construct mass, angular
momentum and entropy of black hole solution of Generalized Minimal Massive Gravity (GMMG)
in asymptotically Anti-de Sitter (AdS) spacetimes. The generalized minimal massive gravity theory is realized
by adding the CS deformation term, the higher derivative deformation term, and an extra term
to pure Einstein gravity with a negative cosmological constant. We apply our result for conserved charge $Q^{\mu} (\bar{\xi})$ to the rotating BTZ black hole solution of GMMG, and find energy, angular momentum and entropy. Then we show our results for these quantities are consistent with the first law of black holes thermodynamics.
\end{abstract}

\newpage

\section{Introduction}
We know that the pure Einstein-–Hilbert gravity in three dimensions exhibits no propagating physical degrees of freedom \cite{2',3'}. But adding the gravitational Chern-Simons term produces a
propagating massive graviton \cite{4'}. The resulting theory
is called topologically massive gravity (TMG). Including a negative cosmological constant, yields cosmological topologically massive
 gravity (CTMG). In this case the theory exhibits both gravitons and black holes. Unfortunately there is a problem in this model, with the usual
sign for the gravitational constant, the massive excitations of CTMG carry negative energy. In the absence of a cosmological constant, one can
change the sign of the gravitational constant, but if $\Lambda <0$, this will give a negative mass to the BTZ
black hole, so the existence of a stable ground state is in doubt in this model \cite{5}. Recently an interesting three dimensional massive gravity introduced by Bergshoeff, et.al \cite{17} which dubbed Minimal Massive Gravity (MMG), which has the same minimal local structure as Topologically Massive Gravity (TMG) \cite{4'}. The MMG model has the same gravitational
degree of freedom as the TMG has and the linearization of the metric field equations for MMG yield a single propagating
massive spin-2 field. So both models have the same spectrum \cite{18}. However, in contrast to TMG, there is not bulk vs boundary clash in the framework of this new model. During last months some interesting works have been done on MMG model \cite{18}. More recently, this model has been extended to General Minimal Massive Gravity theory (GMMG) \cite{1'}. GMMG is a unification of MMG with New Massive Gravity (NMG) \cite{6}, so this model is realized by adding the higher derivative deformation term to the Lagrangian of MMG.\\ In this paper we want to construct mass, angular
momentum and entropy of black hole solution of GMMG
in asymptotically Anti-de Sitter (AdS) spacetimes.
There are several approach to obtain mass and angular momentum of black holes for higher
curvature theories \cite{1}-\cite{12}.
 Arnowit-Deser-Misner (ADM) method \cite{9'} uses a linearization of metric around asymptotically flat spacetime, so this approach fails here because we consider the solution which are not asymptotically flat. A method to calculate the energy of asymptotically AdS solution was given by Abbott and Deser \cite{1}. Deser and Tekin have extended this approach to the calculation of the energy of asymptotically dS or AdS solutions in higher curvature gravity models and also to TMG \cite{3}. In contrast to the ADM method, this ADT formalism is covariant. Another method is the Brown-York formalism \cite{12'} which is based on quasi-local concept, but this approach also is not covariant. The authors of \cite{13'} have obtained the quasi-local conserved charges for  black holes in any diffeomorphically invariant theory of gravity. By considering an appropriate variation of the metric, they have established a one-to-one correspondence between the ADT approach and the linear Noether expressions. They have extended this work to a theory of gravity containing a gravitational Chern-Simons term in \cite{14'}, and have computed the off-shell potential and quasi-local conserved charges of some black holes in TMG. We should mention that before these works, the authors of \cite{15'} have computed the ADT charges for a solution of TMG linearized about an arbitrary background and have applied the result to evaluate the mass and angular momentum of the non-asymptotically flat, non-asymptotically AdS black hole solution (ACL black hole) of TMG. One another way to obtain mass and angular momentum of black holes for higher curvature models in the case of asymptotically AdS space is application of AdS/CFT correspondence, because the definition of conserved charges in the dual field theory is clear and no any ambiguities in their construction \cite{7,9,10,11}. This method is covariant and takes into account the non-linear effects. However this formalism is applicable only to asymptotically (warped) AdS space. Moreover to obtain holographic conserved charges, one needs the boundary stress tensor which depends on the explicit form of Gibbons-Hawking terms \cite{c} and counter term which are not known in general. So this approach becomes complicated for a higher derivative model of gravity \cite{16'}.  Another way is the super angular momentum approach \cite{17',18'}, by this method one can compute the conserved charges with non-linear effects. In this formalism the non-linear conserved charge obtained by the first integral of equation of motion. However this method is not completely covariant, moreover it is inconsistent with the first law of black hole thermodynamics for warped $AdS_3$ black hole solution of TMG.
 Here we follow the method given by Abbott, Deser, and Tekin in \cite{1,2,3}, which need to obtain the field equations and linearize them about the (A)dS vacuum of the model. By this method we obtain conserved charges which are consistent with the first law of black hole thermodynamics.
\\Our paper is organized as follows. In Sec.2 we review GMMG briefly. In Sec.3 we
will obtain the formula for the calculation of conserved charges in this model in asymptotically AdS$_3$ spacetime. Then we apply our result for conserved charge $Q^{\mu} (\bar{\xi})$ to the rotating BTZ black hole solution of GMMG, and find energy, angular momentum and entropy. Sec.4 is devoted to conclusions and discussions.

\section{The Generalized Minimal Massive Gravity}
We introduce the Lagrangian
of GMMG model as \cite{1'}
  \begin{equation} \label{2}
 L_{GMMG}=L_{GMG}+\frac{\alpha}{2}e.h\times h
\end{equation}
where
\begin{equation} \label{2'}
 L_{GMG}=L_{TMG}-\frac{1}{m^2}(f.R+\frac{1}{2}e.f\times f)
\end{equation}
here $m$ is mass parameter of NMG term and $f$ is an auxiliary one-form field. $L_{TMG}$ is the Lagrangian of TMG,
 \begin{equation} \label{1'}
 L_{TMG}=-\sigma e.R+\frac{\Lambda_0}{6} e.e\times e+h.T(\omega)+\frac{1}{2\mu}(\omega.d\omega+\frac{1}{3}\omega.\omega\times\omega)
\end{equation}
 where $\Lambda_0$ is a cosmological parameter with dimension of mass squared, and $\sigma$ a sign. $\mu$ is mass parameter of Lorentz Chern-Simons term. $\alpha$ is a dimensionless parameter, $e$ is dreibein, $h$ is the auxiliary field, $\omega$ is dualised  spin-connection, $T(\omega)$ and
  $R(\omega)$ are Lorentz covariant torsion and curvature 2-form respectively. So by adding extra term $\frac{\alpha}{2}e.h\times h$ to the Lagrangian of generalized massive gravity we obtain Lagrangian of GMMG model. The equation for metric can be obtained by generalizing  field equation of MMG. Due to this we introduce GMMG field equation
as follows \cite{1'}
\begin{equation}\label{8}
\Phi ^{\mu \nu} = \Lambda _{0} g^{\mu \nu} + \sigma G^{\mu \nu} + \frac{1}{\mu} C^{\mu \nu} + \frac{\gamma}{\mu ^{2}}
J^{\mu \nu} + \frac{s}{2 m^{2}} K^{\mu \nu}=0 ,
\end{equation}
where $ G ^{\mu \nu}$ is Einstein's tensor, the Cotton tensor is
\begin{equation}\label{9}
C^{\mu \nu}=\frac{1}{\sqrt{-g}} \varepsilon ^{\mu \alpha \beta} \nabla _{\alpha} S^{\nu}_{\beta},
\end{equation}
where $ S^{\mu}_{\nu}=R^{\mu}_{\nu}-\frac{1}{4}\delta^{\mu}_{\nu}R $ is the Schouten tensor in 3 dimensions,
\begin{equation}\label{10}
J^{\mu \nu} =\frac{1}{2g}  \varepsilon ^{\mu \rho \sigma}  \varepsilon ^{\nu \alpha \beta} S_{\rho \alpha} S_{\sigma \beta},
\end{equation}
and
\begin{equation}\label{11}
K^{\mu \nu}=2 \Box R^{\mu \nu} - \frac{1}{2} \nabla ^{\mu} \nabla ^{\nu} R - \frac{1}{2} g^{\mu \nu} \Box R - 8 R^{\mu \alpha}
R^{\nu}_{\alpha}+\frac{9}{2} R R^{\mu \nu} + 3 g^{\mu \nu}  R^{\alpha \beta} R_{\alpha \beta} - \frac{13}{8} g^{\mu \nu} R^{2} ,
\end{equation}
$s$ is sign, $\gamma$, $\sigma$ $\Lambda_{0}$ are the parameters
 which defined in terms of cosmological constant $\Lambda=\frac{-1}{l^2}$, $m$, $\mu$, and the sign of Einstein-Hilbert term.
Here $G_{mn}$ and $C_{mn}~$denote Einstein tensor and Cotton tensor
respectively. Symmetric tensors $J_{mn}~$and $K_{mn}$ are coming from MMG
and NMG parts respectively \cite{6,17}.
\section{Charges of GMMG}
In this section we would like to obtain the conserved charges of GMMG for asymptotically (A)dS space-times. Here we follow the method given in \cite{1,2,3} (see also \cite{4}), which need to obtain the field equations and linearize them about the (A)dS vacuum of the model. \footnote{Here we should mention that  although we obtain the conserved charges from the linearization of the field equations around a background, but the method is general. There is no dependence on background (except that there must be asymptotic
Killing vectors and spatial infinity of course) in the method of \cite{2}. Multiple vacua are universal features of all $R+R^2$ etc models, since they clearly allow both flat and (A)dS vacua, but energy is still definable around each branch, though one is unstable (see for example \cite{de}).  }\\
The field equations of the model can be written as
\begin{equation}\label{1}
\Phi   _{\mu \nu} (g, R, \nabla (Riemann) , R ^{2} , ....) = 8 \pi T _{\mu \nu}.
\end{equation}
We assume that (A)dS is the background solution $ \Phi  _{\mu \nu} (\bar{g}) = 0$. The linearized form of the above equation can be written symbolically as
\begin{equation}\label{2}
 \mathcal{O} (\bar{g})_{\mu \nu \alpha \beta} h^{\alpha \beta}
 = 8 \pi T _{\mu \nu} ,
\end{equation}
where the deviation of background is, $h_{\mu \nu}=g_{\mu \nu}-\bar{g}_{\mu \nu}$.
If the field equation (\ref{1}) come from an diffeomorphism invariant action, then we have
\begin{equation}\label{3}
\nabla _{\mu} \Phi ^{\mu \nu} = 0,
\end{equation}
From  (\ref{2}) and  (\ref{3}) we have
\begin{equation}\label{4}
\bar{\nabla}_{\mu} \mathcal{O} ^{\mu } \hspace{0.5 mm} _{ \nu \alpha \beta} h^{\alpha \beta}=0,
\end{equation}
In order to define globally conserved charges, we use the killing vector $ \bar{\xi} $, and energy-momentum $ T^{\mu \nu} $, then we can define conserved current as
\begin{equation}\label{5}
\sqrt{-\bar{g}} \bar{\nabla} _{\mu} (\bar{\xi}_{\nu} T^{\mu \nu}) = \partial _{\mu} ( \sqrt{-\bar{g}} \bar{\xi}_{\nu} T^{\mu \nu})=0,
\end{equation}
So we obtain conserved charge by
\begin{equation}\label{6}
Q^{\mu} (\bar{\xi})= c \int _{\mathcal{M}} d^{D-1}x \sqrt{-\bar{g}} \bar{\xi}_{\nu} T^{\mu \nu}
 \equiv c \int _{\Sigma}  dl_{i} \mathcal{F}^{\mu i},
\end{equation}
where we have used Stoke's theorem,
 $c$ is an arbitrary constant, and $\mathcal{M}$ is a $(D-1)$-dimensional spatial manifold with boundary $\Sigma$.
We have assumed that $\bar{\xi}_{\nu} T^{\mu \nu} = \bar{\nabla} _{\nu} \mathcal{F}^{\mu \nu}$, where $\mathcal{F}^{\mu \nu}$ is an anti-symmetric tensor. For the background metric we have:
\begin{equation}\label{7}
\bar{R}_{\mu \alpha \nu \beta} = \Lambda \left( \bar{g}_{\mu \nu} \bar{g}_{\alpha \beta} - \bar{g}_{\mu \beta} \bar{g}_{\nu \alpha} \right) , \hspace{1 cm} \bar{R}_{\mu \nu} = 2 \Lambda \bar{g}_{\mu \nu}, \hspace{1 cm} \bar{R}=6 \Lambda,
\end{equation}
Now we obtain the linearized form of field equation (\ref{8}) around the AdS$_3$ space-time. So at the first order the field equation can be written as
\begin{equation}\label{19}
\Phi ^{\mu \nu} (\bar{g}) + \Phi _{L} ^{\mu \nu} = 8 \pi T ^{\mu \nu} ,
\end{equation}
where
\begin{equation}\label{20}
\Phi _{L}^{\mu \nu} = - \Lambda _{0} h^{\mu \nu} + \sigma G_{L}^{\mu \nu} + \frac{1}{\mu} C_{L}^{\mu \nu}
+ \frac{\gamma}{\mu ^{2}} J_{L}^{\mu \nu} + \frac{s}{2 m^{2}} K_{L}^{\mu \nu} ,
\end{equation}
\begin{equation}\label{14}
\mathcal{G}^{L}_{\mu \nu}= R^{L}_{\mu \nu} - \frac{1}{2}\bar{g}_{\mu \nu} R^{L} - 2 \Lambda h_{\mu \nu}  ,
\end{equation}
\begin{equation}\label{15}
C_{L}^{\mu \nu}=\frac{1}{\sqrt{-\bar{g}}} \varepsilon ^{\mu \alpha \beta} \bar{g}_{\beta \sigma} \nabla _{\alpha}
 \left( R_{L}^{\sigma \nu}  - \frac{1}{4} \bar{g}^{\sigma \nu}  R_{L}+2 \Lambda h^{\sigma \nu} \right) ,
\end{equation}

\begin{equation}\label{17}
J_{L}^{\mu \nu} = -\frac{1}{2} \Lambda \mathcal{G}_{L}^{\mu \nu} -\frac{1}{4} \Lambda ^{2} h^{\mu \nu}
\end{equation}
\begin{equation}\label{18}
K_{L}^{\mu \nu}=2 \bar{\Box} \mathcal{G}_{L}^{\mu \nu} + \frac{1}{2} \bar{g}^{\mu \nu} \bar{\Box} R_{L}
-\frac{1}{2} \bar{\nabla} ^{\mu} \bar{\nabla} ^{\nu} R_{L} - 5 \Lambda \mathcal{G}_{L}^{\mu \nu} - \Lambda \bar{g}^{\mu \nu} R_{L} +\frac{1}{2} \Lambda ^{2} h^{\mu \nu},
\end{equation}
Here we have defined that $ \mathcal{G}_{\mu \nu} = G_{\mu \nu}+ \Lambda g_{\mu \nu} $. The linear form of Ricci tensor and Ricci scalar are given by following equation respectively
\begin{equation}\label{12}
R^{L}_{\mu \nu} = \frac{1}{2} \left( - \bar{\Box} h_{\mu \nu} - \bar{\nabla} _{\mu} \bar{\nabla} _{\nu} h
+\bar{\nabla} ^{\lambda} \bar{\nabla} _{\mu} h_{\lambda \nu} +\bar{\nabla} ^{\lambda} \bar{\nabla} _{\nu} h_{\lambda \mu} \right) ,
\end{equation}
\begin{equation}\label{13}
R^{L}=- \bar{\Box} h + \bar{\nabla}_{\mu} \bar{\nabla} _{\nu} h^{\mu \nu} - 2 \Lambda h  ,
\end{equation}
Using the Ricci tensor and Ricci scalar of AdS$_3$ background in (\ref{7}), it is easy to see that
\begin{equation}\label{21}
\bar{G}^{\mu \nu}=-\Lambda \bar{g}^{\mu \nu} ,\hspace{5 mm} \bar{C}^{\mu \nu}=0,\hspace{5 mm}
 \bar{J}^{\mu \nu}=\frac{1}{4}\Lambda ^{2}\bar{g}^{\mu \nu},\hspace{5 mm} \bar{K}^{\mu \nu}=-\frac{1}{2}\Lambda ^{2}\bar{g}^{\mu \nu},
\end{equation}
Then field equation for $AdS_{3}$ reduces to an quadratic equation for
\begin{equation}\label{22}
\Lambda _{0} - \sigma \Lambda + \frac{\gamma \Lambda ^{2}}{4 \mu ^{2}} - \frac{s \Lambda ^{2}}{4 m^{2}}=0 ,
\end{equation}
so,
\begin{equation}  \label{51}
\Lambda=\frac{({\sigma}\pm\sqrt{{\sigma}^{2}-{\Lambda}_{0}(\frac{\gamma}{\mu^2}-\frac{s}{m^2})})}{\frac{1}{2}
(\frac{\gamma}{\mu^2}-\frac{s}{m^2})}
\end{equation}

Since $\Phi _{\mu \nu}(\bar{g})=0$, (\ref{19}) takes following form
\begin{equation}\label{23}
\Phi _{L} ^{\mu \nu} = 8\pi T _{\mu \nu}.
\end{equation}
Substituting (\ref{14})-(\ref{18}) and (\ref{22}) in (\ref{23}), we have
\begin{equation}\label{24}
\left( \sigma \Lambda - \frac{\gamma \Lambda}{2 \mu ^{2}} \right) \mathcal{G}_{L}^{\mu \nu} + \frac{1}{\mu} C_{L}^{\mu \nu}
+ \frac{s}{2 m^{2}} \left( K_{L}^{\mu \nu} - \frac{1}{2}\Lambda ^{2} h^{\mu \nu} \right) = 8 \pi T ^{\mu \nu} ,
\end{equation}
One can show that $ \bar{\nabla}_{\nu} \mathcal{G}_{L} ^{\mu \nu } = \bar{\nabla}_{\nu} C_{L} ^{\mu \nu } = 0$, then
by the following identities \cite{3}
$$ \bar{\nabla}_{\nu} \left[ (  \bar{g}^{\mu \nu} \bar{\Box} - \bar{\nabla}^{\mu} \bar{\nabla}^{\nu}
+ 2 \Lambda  \bar{g}^{\mu \nu} ) R_{L} \right] =0, $$
\begin{equation}\label{25}
\bar{\nabla}_{\nu} \left[ \bar{\Box} \mathcal{G}_{L} ^{\mu \nu } -\Lambda \bar{g}^{\mu \nu} R_{L} \right] =0,
\end{equation}
 we conclude that
 \begin{equation}\label{26}
\bar{\nabla}_{\nu} \left( K_{L}^{\mu \nu} - \frac{1}{2}\Lambda ^{2} h^{\mu \nu} \right) = 0.
\end{equation}
So (\ref{24}) obeys the Bianchi identities, and we can use (\ref{24}) for definition of conserved charges.
  One can check that \cite{3, 13}
 $$ \sqrt{-\bar{g}} \bar{\xi}_{\nu} \mathcal{G}_{L} ^{\mu \nu } =\frac{1}{2} \partial _{\nu} q_{E}^{\mu \nu} (\bar{\xi}) ,\hspace{1 cm}
\sqrt{-\bar{g}} \bar{\xi}_{\nu} C_{L} ^{\mu \nu } =\frac{1}{2} \partial _{\nu} \left(  \frac{1}{2 \mu} q_{E}^{\mu \nu} (\bar{\Xi})
+  \frac{1}{2 \mu} q_{C}^{\mu \nu} (\bar{\xi}) \right), $$
\begin{equation}\label{27}
\sqrt{-\bar{g}} \bar{\xi}_{\nu} \left( K_{L}^{\mu \nu} - \frac{1}{2}\Lambda ^{2} h^{\mu \nu} \right) =
 \frac{1}{2} \partial _{\nu} \left( q_{N}^{\mu \nu} (\bar{\xi}) -\Lambda q_{E}^{\mu \nu} (\bar{\xi}) \right)  ,
\end{equation}
 where
 $$ q_{E}^{\mu \nu} (\bar{\xi}) = 2 \sqrt{-\bar{g}} \left( \bar{\xi}_{\lambda}  \bar{\nabla}^{ [ \mu} h^{\nu ] \lambda}
 + \bar{\xi}^{ [ \mu}  \bar{\nabla}^{\nu ] } h + h^{\lambda [ \mu}  \bar{\nabla}^{\nu ] } \bar{\xi}_{\lambda} +
\bar{\xi} ^{ [ \nu} \bar{\nabla} _{ \lambda} h^{\mu ] \lambda} + \frac{1}{2} h \bar{\nabla} ^{ \mu} \bar{\xi}^{\nu} \right) , $$
$$ q_{C}^{\mu \nu} (\bar{\xi}) = \varepsilon ^{\mu \nu}  _{\hspace{3 mm} \alpha} \mathcal{G}_{L}^{\alpha \beta} \bar{\xi}_{\beta}
 + \varepsilon ^{\beta \nu} _{\hspace{3 mm} \alpha} \mathcal{G}_{L}^{\mu \alpha} \bar{\xi}_{\beta}
 +\varepsilon ^{\mu \beta}  _{\hspace{3 mm} \alpha} \mathcal{G}_{L}^{\alpha \nu} \bar{\xi}_{\beta}, $$
 \begin{equation}\label{28}
q_{N}^{\mu \nu} (\bar{\xi}) = \sqrt{-\bar{g}} \left[ 4 \left(  \bar{\xi}_{\lambda}  \bar{\nabla}^{ [ \nu} \mathcal{G}_{L}^{\mu ] \lambda}
+\mathcal{G}_{L}^{\lambda [ \nu} \bar{\nabla} ^{ \mu ] } \bar{\xi}_{\lambda} \right) + \bar{\xi}^{ [ \mu}  \bar{\nabla}^{\nu ] } R_{L}
+ \frac{1}{2} R_{L} \bar{\nabla} ^{ \mu} \bar{\xi}^{\nu} \right] ,
\end{equation}
also, $ \bar{\Xi}^{\beta}=\frac{1}{\sqrt{-\bar{g}}} \varepsilon^{\alpha \lambda \beta} \bar{\nabla} _{ \alpha} \bar{\xi}^{\lambda}  $.
Using (\ref{27}), we can rewrite (\ref{24}) as
 \begin{equation}\label{29}
16 \pi \sqrt{-\bar{g}} \bar{\xi}_{\nu} T^{\mu \nu}=\partial_{\nu} \left[ \left( \sigma -\frac{\gamma \Lambda}{2 \mu ^{2}} - \frac{s \Lambda}{2m^{2}} \right) q_{E}^{\mu \nu} (\bar{\xi})+\frac{1}{2 \mu} q_{E}^{\mu \nu} (\bar{\Xi}) + \frac{1}{2 \mu} q_{C}^{\mu \nu} (\bar{\xi}) +\frac{s}{2m^{2}} q_{N}^{\mu \nu} (\bar{\xi}) \right] ,
\end{equation}
substituting this result in (\ref{6}), we obtain
\begin{equation}\label{30}
Q^{\mu} (\bar{\xi})=\frac{c}{16 \pi} \int _{\Sigma}  dl_{i} \left[ \left( \sigma -\frac{\gamma \Lambda}{2 \mu ^{2}} - \frac{s \Lambda}{2m^{2}} \right) q_{E}^{\mu i} (\bar{\xi})+\frac{1}{2 \mu} q_{E}^{\mu i} (\bar{\Xi}) + \frac{1}{2 \mu} q_{C}^{\mu i} (\bar{\xi}) +\frac{s}{2m^{2}} q_{N}^{\mu i} (\bar{\xi}) \right],
\end{equation}
 where $i$ denotes the space direction orthogonal to the boundary $\Sigma$.\\
 In the limiting case $\frac{1}{m^{^2}}\rightarrow 0$, where GMMG reduce to the MMG, our result for conserved charge $Q^{\mu} (\bar{\xi})$ in above equation reduce to the result of \cite{13} for MMG. Now we apply Eq.(\ref{30}) to the rotating BTZ black hole solution of GMMG, in order to obtain the energy, angular momentum and entropy of this black hole. The BTZ line-element is
 \begin{equation}\label{31}
ds^{2}=-\frac{(r^{2}-r_{+}^{2})(r^{2}-r_{-}^{2})}{l^{2} r^{2}} dt^{2} +
 \frac{l^{2} r^{2}}{(r^{2}-r_{+}^{2})(r^{2}-r_{-}^{2})}dr^{2} + r^{2} \left( d\phi - \frac{r_{+}r_{-}}{lr^{2}}dt \right)^{2} ,
\end{equation}
where $\Lambda=-\dfrac{1}{l^2}$, also $r_{+}$ and $r_{-}$ are outer and inner horizon respectively. The case of $r_{+}=r_{-}=0$ is correspond to the  background. Then, one can read that
\begin{equation}\label{32}
h_{tt}=\frac{r_{+}^{2}+r_{-}^{2}}{l^2},\hspace{5 mm} h_{t\phi}=-\frac{r_{+}r_{-}}{l},\hspace{5 mm}
 h_{rr}=\frac{l^2(r_{+}^{2}+r_{-}^{2})}{r^4},
\end{equation}
$\Sigma$ is a circle and $dl_{i}=(d\phi , 0) $.
\\ Energy is correspond to the Killing vector $\bar{\xi}=\partial _{t}$ and $c=-8$ then
\begin{equation}\label{33}
E=\left( \sigma + \frac{\gamma}{2\mu ^{2}l^{2}} + \frac{s}{2 m^{2} l^{2}} \right) \frac{r_{+}^{2}+r_{-}^{2}}{l^2}
- \frac{2 r_{+}r_{-}}{\mu l^{3}},
\end{equation}
In the other hand angular momentum of BTZ black hole is correspond to the Killing vector $\bar{\xi}=\partial _{\phi}$ and $c=8$, so we have
\begin{equation}\label{34}
J=\left( \sigma + \frac{\gamma}{2\mu ^{2}l^{2}} + \frac{s}{2 m^{2} l^{2}} \right) \frac{2 r_{+}r_{-}}{l}
- \frac{r_{+}^{2}+r_{-}^{2}}{\mu l^2},
\end{equation}
If one write the metric of rotating BTZ black hole in terms of mass $M$ and angular momentum parameter $a$, the above expression for energy $E$ and angular momentum $J$  can be rewritten as
\begin{equation}\label{330}
E=\left[( \sigma-\frac{\gamma\Lambda}{2\mu^{2}}-\frac{s\Lambda}{2m^{2}})M+\frac{\Lambda a}{\mu}  \right]
\end{equation}
 \begin{equation}\label{340}
J=\left[( \sigma-\frac{\gamma\Lambda}{2\mu^{2}}-\frac{s\Lambda}{2m^{2}})a-\frac{M}{\mu} \right]
\end{equation}
The above equations reduce to the corresponding results for MMG in the limit $\frac{1}{m^{^2}}\rightarrow 0$, \cite{13}.
If we take $\bar{\xi}=\partial _{t} + \frac{r_{-}}{l r_{+}}\partial _{\phi}$ and $c=-\frac{32 \pi}{\kappa} $,
where $\kappa =\frac{r_{+}^{2}-r_{-}^{2}}{l^2 r_{+}} $ is surface gravity, then
\begin{equation}\label{35}
S=4 \pi \left[ \left( \sigma + \frac{\gamma}{2\mu ^{2}l^{2}} + \frac{s}{2 m^{2} l^{2}} \right) r_{+} - \frac{r_{-}}{\mu l} \right] ,
\end{equation}
one can check that these results satisfy the first law of thermodynamics,that is
\begin{equation}\label{36}
dE=T_{H}dS+\Omega _{H} dJ,
\end{equation}
where  $T_{H}=\frac{\kappa}{2 \pi} $ and $ \Omega _{H} = \frac{r_{-}}{l r_{+}} $.

\section{Conclusion}
In this paper we have investigated the Abbott-Deser-Tekin charge construction in the framework of Generalized Minimal Massive Gravity  in asymptotically AdS space-time. We have applied our result for conserved charge $Q^{\mu} (\bar{\xi})$  in Eq.(\ref{30}) to the rotating BTZ black hole solution of GMMG. In the limit $\frac{1}{m^{^2}}\rightarrow 0$, where GMMG reduce to the MMG model, our conserved charge  $Q^{\mu} (\bar{\xi})$ reduce to the corresponding result for MMG, which has been obtained in \cite{13}. By this method and correspond to the Killing vector fields $\bar{\xi}=\partial _{t}$ and $\bar{\xi}=\partial _{\phi}$, we have obtained energy and angular momentum of rotating BTZ  black hole respectively. After that by considering Killing vector field $\bar{\xi}=\partial _{t} + \frac{r_{-}}{l r_{+}}\partial _{\phi}$  we have obtained the entropy of BTZ black hole. Then we have shown that our result for entropy is consistent with the first law of black holes thermodynamics.
\section{Acknowledgments}
M. R. Setare  grateful to  S. Deser for helpful discussions and correspondence.

\end{document}